\documentclass[aps,pra,notitlepage,superscriptaddress,amsmath,amssymb]{revtex4-1}
\usepackage{graphicx}
\pdfoutput=1

\begin{document}

\title{Nonlinear quantum optics for spinor slow light}

\author{J. Ruseckas}
\affiliation{Institute of Theoretical Physics and Astronomy, Vilnius University,
Saul\.etekio 3, LT-10257 Vilnius, Lithuania}
\email{julius.ruseckas@tfai.vu.lt}
\homepage{http://web.vu.lt/tfai/j.ruseckas}

\author{V. Kudria\v{s}ov}
\affiliation{Institute of Theoretical Physics and Astronomy, Vilnius University,
Saul\.etekio 3, LT-10257 Vilnius, Lithuania}

\author{A. Mekys}
\affiliation{Institute of Theoretical Physics and Astronomy, Vilnius University,
Saul\.etekio 3, LT-10257 Vilnius, Lithuania}

\author{T. Andrijauskas}
\affiliation{Institute of Theoretical Physics and Astronomy, Vilnius University,
Saul\.etekio 3, LT-10257 Vilnius, Lithuania}

\author{Ite A. Yu}
\affiliation{Department of Physics, National Tsing Hua University, Hsinchu
30013, Taiwan}
\affiliation{Center for Quantum Technology, Hsinchu 30013, Taiwan}

\author{G. Juzeli\=unas}
\affiliation{Institute of Theoretical Physics and Astronomy, Vilnius University,
Saul\.etekio 3, LT-10257 Vilnius, Lithuania}

\date{\today{}}

\begin{abstract}
We investigate quantum nonlinear effects at a level of individual quanta in a
double tripod atom-light coupling scheme involving two atomic Rydberg states.
In such a scheme the slow light coherently coupled to strongly interacting
Rydberg states represents a two-component or spinor light.  The scheme provides
additional possibilities for the control and manipulation of light quanta.  A
distinctive feature of the proposed setup is that it combines the spin-orbit
coupling for the spinor slow light with an interaction between the photons,
enabling generation of the second probe beam even when two-photon detuning is
zero.  Furthermore, the interaction between the photons can become repulsive if
the one-photon detunings have opposite signs. This is different from a single
ladder atom-light coupling scheme, in which the interaction between the photons
is attractive for both positive and negative detunings, as long as the Rabi
frequency of the control beam is not too large.
\end{abstract}
\maketitle

\section{Introduction}

Atoms excited to high-lying Rydberg states with a principal quantum number $n$
above 50 have recently attracted a significant attention \cite{Saffman2010}.
Since the van der Waals interaction between atoms increases with the principal
quantum number as $n^{11}$ , the interaction between the Rydberg atoms is
enhanced by many orders of magnitude compared to the interaction between atoms
in the ground state \cite{Bohlouli2007,Beguin2013}. The strong interaction
between the Rydberg atoms prevents a simultaneous excitation of nearby atoms
leading to the Rydberg blockade \cite{Tong2004,Singer2004,Heidemann2007}.
These properties of the Rydberg atoms have found application in the quantum
information processing
\cite{Jaksch2000,Lukin2001,Gaeetan2009,Urban2009,Isenhower2010}, studies of
interacting many-body systems
\cite{Cinti2010,Pohl2010,Pupillo2010,Viteau2011,Carr2013,Schauss2015}, as well
as non-linear quantum optics for slow light
\cite{Petrosyan2011,Gorshkov2011,Dudin2012,Dudin2012a,Peyronel2012,Firstenberg2013,Li2013,Chang2014}.
The latter application is based on the fact that the Rydberg interaction brings
neighboring Rydberg atoms out of the resonance destroying the
electromagnetically induced transparency (EIT). Consequently the closeby
Rydberg atoms absorb the slow light which becomes anti-bunched during its
propagation through the atomic medium
\cite{Petrosyan2011,Gorshkov2011,Dudin2012,Dudin2012a,Peyronel2012,Firstenberg2013,Li2013,Chang2014}.
Dispersive coupling of light to strongly interacting atoms in highly excited
Rydberg states allows to induce an effective attractive force between the
propagating Rydberg polaritons and to create bound states of the polaritons
\cite{Firstenberg2013,Bienias2014}. For a recent review of nonlinear quantum
optics mediated by Rydberg interactions see Ref.~\cite{Firstenberg2016}.

In a usual Rydberg EIT, a ladder atom-light coupling configuration is typically
employed involving an atomic ground state, an intermediate excited state and a
Rydberg state \cite{Gorshkov2011,Peyronel2012}.  In the present paper we study
a more complicated double tripod level scheme for the Rydberg EIT, providing
new physical effects not present in the ladder scheme.  The double tripod
scheme involves two probe laser fields and there are two adiabatic eigenmodes
propagating inside the atomic medium that are imune to spontaneous decay. They
form a two-component (spinor) slow light.  In the case of non-interacting atoms
the propagation of the two-component slow light was considered in
Refs.~\cite{Unanyan2010,Ruseckas2011,Ruseckas2013} and was experimentally
demonstrated recently \cite{Lee2014}. The double tripod atom-light coupling
scheme provides a possibility to create the spin-orbit coupling for the spinor
slow light. The influence of the spin-orbit coupling on the propagation of
light have been investigated in several studies
\cite{Berard2004,Bliokh2004,Bliokh2004a,Onoda2004,Berard2006} leading to a
generalization of geometric optics called geometric spintronics
\cite{Duval2006,Duval2007}. The present study adds the interaction effects to
the spinor slow light.

Here we include interactions between the atoms excited to the Rydberg levels
leading to non-linear effects for the spinor slow light. In contrast to a
ladder scheme where a single one-photon detuning can be present, the probe
beams in the double tripod setup can have different one-photon detunings. We
show that the propagation of the slow light significantly depend not only on
the magnitude of one-photon detunings, but also on the their difference. For
large and equal one-photon detunings the propagation of the spinor slow light
combines a spin-orbit coupling at the distances between photons larger than the
Rydberg blockade radius together with an effective attractive interaction of
photons at small distances. The latter attractive force is the same as in a
ladder scheme used in Ref.~\cite{Firstenberg2013}. Yet, if the one-photon
detunings have opposite signs, the photons from the different probe beams
experience repulsion instead of attraction.

Different from a simple ladder scheme, in the double-tripod setup the photons
can be transferred from one probe beam to another.  Although such a transfer is
possible in a double-tripod setup without atom-atom interactions, this can be
accomplished only for a nonzero two-photon detuning
\cite{Ruseckas2013,Lee2014}.  In this paper we show that the interactions
between Rydberg atoms together with the spin-orbit coupling makes it possible
the second probe beam to appear even when two-photon detuning is zero and the
envelope of the input probe field is constant most of the time.  

Note, that some of the phenomena considered here can be explored in
simpler setups. For example, conversion between the two probe beams can be
achieved in double-Lambda linkage pattern involving only one Rydberg level.
However, in the double tripod setup there are two adiabatic  eigenmodes
immune to spontaneous decay, whereas in the double-Lambda setup there is only
one. Two adiabatic  eigenmodes are needed to create a slow ligtht polariton
that behaves as a spinor-like object and is subjected to a spin-orbit
coupling.  To achieve this, simpler settings are not sufficient.

The proposed double-tripod scheme with Rydberg levels could be a suitable
system to explore many-body effects involving the spin-orbit coupling.  The
setup can also lead to novel applications in quantum information
manipulation and nonlinear optics. For example, this scheme can be used to
realize the quantum gates for two-color qubits, where two-color qubit
represents a superposition state of two frequency modes.  Conversion between
different probe photons caused by Rydberg interactions can be utilized for
generation of a second photon of different frequency entangled with the photon
incident on the medium.

The paper is organized as follows. In Sec.~\ref{sec:formulation} we present the
proposed setup and in Sec.~\ref{sec:derivation} we introduce a description of
the system. In Sec.~\ref{sec:equal-detunings} we consider the case when both
one-photon detunings are equal and derive an approximate closed propagation
equation for the two-photon wave function. In Sec.~\ref{sec:opposite-signs} we
investigate the consequences of one-photon detunings having the opposite signs.
Section \ref{sec:conclusions} summarizes the findings.

\section{Formulation\label{sec:formulation}}

\subsection{Atom-light system}

\begin{figure}
\includegraphics[width=0.4\textwidth]{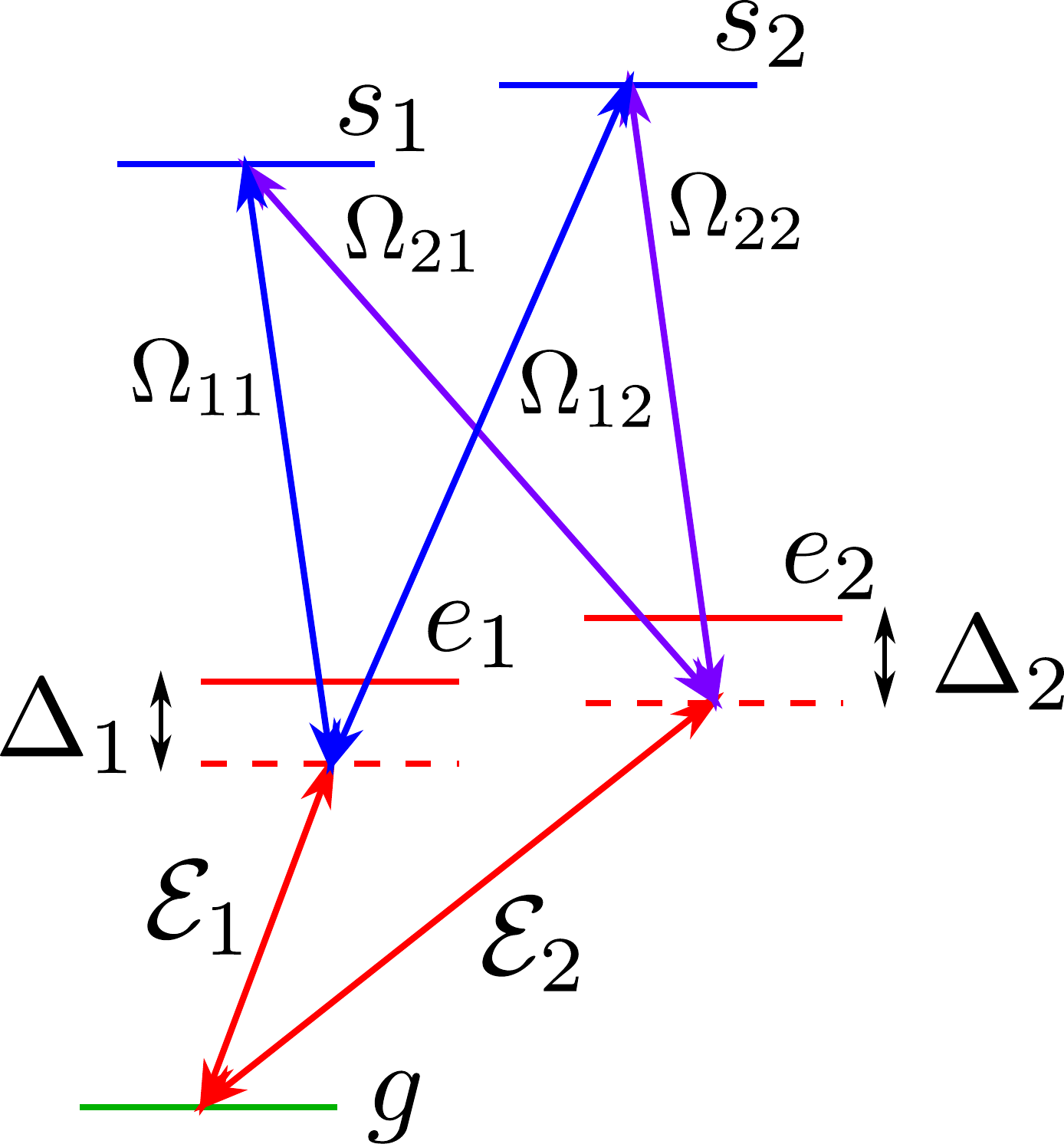}
\caption{Double tripod atom-light coupling scheme involving the ground level
  $g$, the Rydberg levels $s_{1}$ and $s_{2}$, and the intermediate excited
  levels $e_{1}$ and $e_{2}$. Two weak probe beams with the amplitudes
  $\mathcal{E}_{j}$ connect the ground level $g$ to the intermediate levels
  $e_{j}$. Four strong control beam with the Rabi frequencies $\Omega_{jl}$
  connect the intermediate levels $e_{j}$ to the Rydberg levels $s_{j}$.}
\label{fig:double-tripod}
\end{figure}

Let us consider an ensemble of atoms characterized by a double tripod scheme of
energy levels shown in Fig.~\ref{fig:double-tripod}. The scheme includes an
atomic ground level $g$, Rydberg levels $s_{1}$, $s_{2}$, and intermediate
excited levels $e_{1}$, $e_{2}$, the corresponding energies being
$\hbar\omega_{g}$, $\hbar\omega_{s_{j}}$ and $\hbar\omega_{e_{j}}$ ($j=1,2$).
We assume that the levels $s_1$ and $s_2$ are both Rydberg $s$ or $d$ states.
The atoms interact with two probe laser fields of lower intensities and four
control fields of much higher intensities. The probe fields with central
frequencies $\omega_{p_{j}}$ drive the atomic transitions $g\rightarrow e_{j}$.
The control fields characterized by the frequencies $\omega_{c_{j,l}}$ couple
the atomic transitions $e_{j}\rightarrow s_{l}$ ($j=1,2$; $l=1,2$) with the
coupling strengths characterized by the Rabi frequencies $\Omega_{j,l}$.

To simplify the description we consider a narrow cigar shape atom cloud. As in
the previously considered latter scheme \cite{Peyronel2012,Firstenberg2013},
the atomic density is assumed to be constant across the probe beams and the
Rydberg blockade radius is larger than the waists of the beams.  In such a
situation a one-dimensional approximation is suitable for the description of
the propagating probe fields. In the continuum approximation
\cite{Fleischhauer2002,Gorshkov2011} the probe fields and atomic excitations
can be represented by slowly varying operators
$\hat{\mathcal{E}}_{j}^{\dag}(z)$, $\hat{\Psi}_{e_{j}}^{\dag}(z)$ and
$\hat{\Psi}_{s_{j}}^{\dag}(z)$. The former $\hat{\mathcal{E}}_{j}^{\dag}(z)$
describes the creation at the position $z$ of a photon with the frequency
$\omega_{p_{j}}$. The latter material operators $\hat{\Psi}_{e_{j}}^{\dag}(z)$
and $\hat{\Psi}_{s_{j}}^{\dag}(z)$ describe excitation of the intermediate
electronic level $e_{j}$ and the Rydberg level $s_{j}$, respectively. In the
Heisenberg representation, the field operators obey the equal-time commutation
relations
\begin{equation}
[\hat{\mathcal{E}}_{j}(z,t),\hat{\mathcal{E}}_{j}^{\dag}(z',t)]=
[\hat{\Psi}_{e_{j}}(z,t),\hat{\Psi}_{e_{j}}^{\dag}(z',t)]=
[\hat{\Psi}_{s_{j}}(z,t),\hat{\Psi}_{s_{j}}^{\dag}(z',t)]=\delta(z-z')\,,
\label{eq:comm}
\end{equation}
all other equal time commutators being zero. 

Outside the medium the Heisenberg equation of motion for the slowly varying
probe field reads
\begin{equation}
\partial_{t}\hat{\mathcal{E}}_{j}(z,t)=-c\partial_{z}\hat{\mathcal{E}}_{j}(z,t)\,.
\end{equation}
Inside the medium the Heisenberg equations of motion for the annihilation
operators have a form similar to the ones previously considered for three level
$\Lambda$ or ladder schemes \cite{Fleischhauer2002,Gorshkov2007,Gorshkov2011}
\begin{align}
\partial_{t}\hat{\mathcal{E}}_{j}(z,t) & =-c\partial_{z}\hat{\mathcal{E}}_{j}(z,t)
+i\frac{g}{2}\hat{\Psi}_{e_{j}}(z,t)\,,\label{eq:e}\\
\partial_{t}\hat{\Psi}_{e_{j}}(z,t) & =
-\frac{i}{2}\tilde{\Delta}_{j}\hat{\Psi}_{e_{j}}(z,t)
+i\frac{g}{2}\hat{\mathcal{E}}_{j}(z,t)
+\frac{i}{2}\sum_{l}\Omega_{jl}\hat{\Psi}_{s_{l}}(z,t)\,,\label{eq:ex}\\
\partial_{t}\hat{\Psi}_{s_{j}}(z,t) & =
\frac{i}{2}\sum_{l}\Omega_{l,j}^{*}\hat{\Psi}_{e_{l}}(z,t)
-i\hat{U}(z,t)\hat{\Psi}_{s_{j}}(z,t)\,.\label{eq:s}
\end{align}
with 
\begin{equation}
\tilde{\Delta}_{j}=2\Delta_{j}-i\Gamma_{j}\,.\label{eq:Delta_j-tilde}
\end{equation}
To include decay of excited states $e_{j}$ with rates $\Gamma_{j}$, we have
added an imaginary part to the one photon detunings
$\Delta_{j}=\omega_{e_{j}}-\omega_{g}-\omega_{p_{j}}$ replacing the latter by
the complex-valued quantities $\tilde{\Delta}_{j}$ featured in
Eq.~(\ref{eq:ex}). In the following for simplicity we will assume that
$\Gamma_{j}=\Gamma$ is the same for both levels.  In Eq.~(\ref{eq:s}) we have
neglected the decay of a coherence between the ground $g$ and Rydberg $s_{j}$
states. In Eqs.~(\ref{eq:ex})--(\ref{eq:s}) the two-photon detunings
are taken to be zero,
$\omega_{g}+\omega_{p_{j}}+\omega_{c_{j,l}}-\omega_{s_{l}}=0$, yet the
one-photon detunings $\Delta_{j}$ are generally nonzero.  In Eqs.~(\ref{eq:e})
and (\ref{eq:ex}) the parameter $g$ characterizes the atom-photon coupling
strength and is assumed the same for both probe fields. The coupling strength
$g$ is related to the optical density $\alpha$ of the medium via the equation
$g^{2}=c\Gamma\alpha/L$, where $L$ is the length of the medium which extends
from $z=0$ to $z=L$. Finally, the operator 
\begin{equation}
\hat{U}(z,t)=\int dz'\,V(z-z')\sum_{j}\hat{\Psi}_{s_{j}}^{\dag}(z',t)
\hat{\Psi}_{s_{j}}(z',t)\label{eq:interaction-op}
\end{equation}
describes the interaction between the Rydberg atoms. This expression for the
operator $\hat{U}(z,t)$ represents only the density-density interaction between
the Rydberg states, corresponding to the van der Waals interaction.  In
general, the exchange interaction between the Rydberg states can also be
present \cite{Li2014,Gorniaczyk2016}. The exchange interaction strength is
significant only when the difference between the principal quantum numbers is
equal to 1 and becomes much smaller for larger differences \cite{Bijnen2013}.
In the present study we assume that the difference between the principal
quantum numbers of the Rydberg levels is sufficiently large so that the
exchange interaction can be neglected.  Similarly to the exchange interaction,
the energy exchange between atoms can also occur due to the resonant
dipole-dipole interaction (RDDI).  Yet the latter RDDI does not show up when
both Rydberg states are the $s$ or $d$ states  due to the vanishing interaction
matrix element, as it is the case in the situation considered here. For
simplicity the position-dependent interaction strength $V(z)=C_{6}/z^{6}$ is
taken to be the same for the atoms in both Rydberg levels $s_{1}$ and $s_{2}$
in Eq.~(\ref{eq:interaction-op}). Such an assumption is justified when the
difference between the principal quantum numbers of both Rydberg levels is not
too large. This and the previously mentioned restrictions can co-exisit e.g.\
by taking a difference in the principal quantum numbers to be around 3 to 5.
The interaction potential $V(r)$ defines the Rydberg blockade radius
$r_{\mathrm{b}}$ via the equality 
\begin{equation}
V(r_{\mathrm{b}})=\Omega^{2}/\Gamma\,,\label{eq:rb}
\end{equation}
where $\Omega$ is the maximum Rabi frequency of the control beams.  Note that
in Eqs.~(\ref{eq:e})--(\ref{eq:s}) we have omitted the Langevin noise
due to the spontaneous emission \cite{Scully1997}, because it is not important
for our analysis involving a low number of excitations \cite{Gorshkov2013}.

\subsection{Two excitation wave-functions}

Similar to the previously considered propagation of a single probe beam
\cite{Peyronel2012,Firstenberg2013}, the probe fields are assumed to be
sufficiently weak at the input, so that the contribution due to more than two
photons is not important. The initial state $|\Phi\rangle$ of the the quantized
radiation field and atomic excitations can then be written only in terms of
components containing at most two photons.  Envelopes of input probe pulses are
assumed to be long and flat. Thus, except for a short transient period, the
envelopes can be considered to be constant most of the time.

The temporal evolution of the Heisenberg operators (\ref{eq:e})--(\ref{eq:s})
preserves the total number of photons and atomic excitations. The radiation
fields and atomic excitations originating from a single photon input are not
subjected to the atom-atom interactions and propagate through the EIT medium
without a decay. In the single-particle picture, the propagation of the
two-component (spinor) slow light in the double tripod system has been
investigated theoretically \cite{Unanyan2010,Ruseckas2011,Ruseckas2013} and
experimentally \cite{Lee2014}. It has been shown that propagation of the slow
light in the double tripod system causes oscillations between the two probe
fields inside the medium. The oscillations can be described by introducing a
matrix of the group velocity with the matrix elements given by
\cite{Ruseckas2013}
\begin{equation}
v_{j,l}=\frac{c}{g^{2}}\sum_{m}\Omega_{j,m}\Omega_{l,m}^{*}\,.\label{eq:v-matr}
\end{equation}
Note, that when the interactions between atoms are present, the condition for
the EIT can be violated. However, we will use the group velocty matrix
(\ref{eq:v-matr}) as a convenient quantity describing the evolution of the
two-excitation state.

Here we are interested in the two-excitation wave functions \cite{Hafezi2012}
for the double tripod scheme. They are defined as
\begin{eqnarray}
\Phi_{\mathcal{E}_{j}\mathcal{E}_{l}}(z,z',t) & = &
\langle\mathrm{vac}|\hat{\mathcal{E}}_{j}(z,t)\hat{\mathcal{E}}_{l}(z',t)|\Phi\rangle\,,
\label{eq:Phi_EE}\\
\Phi_{\mathcal{E}_{j}s_{l}}(z,z',t) & = &
\langle\mathrm{vac}|\hat{\mathcal{E}}_{j}(z,t)\hat{\Psi}_{s_{l}}(z',t)|\Phi\rangle\,,
\label{eq:Phi_Es}\\
\Phi_{s_{j}s_{l}}(z,z',t) & = &
\langle\mathrm{vac}|\hat{\Psi}_{s_{j}}(z,t)\hat{\Psi}_{s_{l}}(z',t)|\Phi\rangle\,,
\label{eq:Phi_ss}
\end{eqnarray}
where $|\Phi\rangle$ is an initial state of the system comprising the quantized
radiation field and atomic excitations. Since the total number of excitations
is preserved during the temporal evolution, only the two photon part of the
initial state-vector $|\Phi\rangle$ contributes to the wave-functions
(\ref{eq:Phi_EE})--(\ref{eq:Phi_ss}).  Using Eqs.~(\ref{eq:e})--(\ref{eq:s}),
in the following Section we will obtain the equations governing the
two-excitation wave functions (\ref{eq:Phi_EE})--(\ref{eq:Phi_ss}) inside the
medium.

\section{Approximate equations for two-excitation wave functions\label{sec:derivation}}

In this section we will obtain an approximate equations describing the
propagation of the probe fields inside the medium. To simplify the equations we
eliminate the intermediate electronic excited levels $e_{j}$ from the
description, assuming that the decay rate $\Gamma$ or the one-photon detuning
$\Delta_{j}$ are large compared to $g$ or $\Omega_{j,l}$. Adiabatic elimination
of an intermediate level to describe the propagation of Rydberg polariton in a
ladder scheme has been used in Ref.~\cite{Gorshkov2011}. We expand all
operators in the power series of $\tilde{\Delta}_{j}^{-1}$ powers:
$\hat{\mathcal{X}}=\hat{\mathcal{X}}^{(0)}+\hat{\mathcal{X}}^{(1)}+\cdots$ ,
subject to the initial condition
$\hat{\mathcal{X}}(z,0)=\hat{\mathcal{X}}^{(0)}(z,0)$ at $t=0$. Here
$\hat{\mathcal{X}}$ stands for the operators $\hat{\mathcal{E}}_{j}$,
$\hat{\Psi}_{e_{j}}$ , $\hat{\Psi}_{s_{j}}$. Collecting the terms of the same
size in Eq.~(\ref{eq:ex}) we obtain that the operator
$\hat{\Psi}_{e_{j}}^{(0)}$ obeys the equation
\begin{equation}
\partial_{t}\hat{\Psi}_{e_{j}}^{(0)}=-\frac{i}{2}\tilde{\Delta}_{j}
\hat{\Psi}_{e_{j}}^{(0)}
\end{equation}
and thus rapidly changes in time:
$\hat{\Psi}_{e_{j}}^{(0)}(t)=\hat{\Psi}_{e_{j}}(0)\exp(-i\tilde{\Delta}_{j}t/2)$.
The first-order terms in Eq.~(\ref{eq:ex}) lead to the following expression for
the slowly changing part $\hat{\Psi}_{e_{j}}^{(1)}$:
\begin{equation}
\hat{\Psi}_{e_{j}}^{(1)}=\tilde{\Delta}_{j}^{-1}g\hat{\mathcal{E}}_{j}^{(0)}
+\tilde{\Delta}_{j}^{-1}\sum_{l}\Omega_{j,l}\hat{\Psi}_{s_{l}}^{(0)}\,.
\end{equation}
Inserting this expression into Eqs.~(\ref{eq:e}), (\ref{eq:s}) we get the
equations for the slowly changing operators
\begin{align}
\partial_{t}\hat{\mathcal{E}}_{j} & =-c\partial_{z}\hat{\mathcal{E}}_{j}
+i\frac{g^{2}}{2}\tilde{\Delta}_{j}^{-1}\hat{\mathcal{E}}_{j}
+i\frac{g}{2}\tilde{\Delta}_{j}^{-1}\sum_{l}\Omega_{j,l}\hat{\Psi}_{s_{l}}\,,
\label{eq:e-2}\\
\partial_{t}\hat{\Psi}_{s_{j}} & =
i\frac{g}{2}\sum_{l}\tilde{\Delta}_{l}^{-1}\Omega_{l,j}^{*}\hat{\mathcal{E}}_{l}
+\frac{i}{2}\sum_{l,m}\Omega_{l,j}^{*}\tilde{\Delta}_{l}^{-1}\Omega_{l,m}
\hat{\Psi}_{s_{m}}-i\hat{U}\hat{\Psi}_{s_{j}}\,,
\label{eq:s-2}
\end{align}
where we have dropped the upper index $(0)$ for a brevity. Using the
definitions of the two-excitation wave functions
(\ref{eq:Phi_EE})--(\ref{eq:Phi_ss}) and Eqs.~(\ref{eq:e-2}), (\ref{eq:s-2}) we
obtain the following propagation equations:
\begin{align}
\partial_{t}\Phi_{\mathcal{E}_{j}\mathcal{E}_{l}} = &
-c(\partial_{z}+\partial_{z'})\Phi_{\mathcal{E}_{j}\mathcal{E}_{l}}
+i\frac{g^{2}}{2}(\tilde{\Delta}_{j}^{-1}+\tilde{\Delta}_{l}^{-1})
\Phi_{\mathcal{E}_{j}\mathcal{E}_{l}}\nonumber\\
&+i\frac{g}{2}\sum_{m}(\tilde{\Delta}_{l}^{-1}\Omega_{l,m}\Phi_{\mathcal{E}_{j}s_{m}}
+\tilde{\Delta}_{j}^{-1}\Omega_{j,m}\Phi_{s_{m}\mathcal{E}_{l}})\,,
\label{eq:ee-t}\\
\partial_{t}\Phi_{\mathcal{E}_{j}s_{l}} = & -c\partial_{z}\Phi_{\mathcal{E}_{j}s_{l}}
+i\frac{g^{2}}{2}\tilde{\Delta}_{j}^{-1}\Phi_{\mathcal{E}_{j}s_{l}}
+\frac{i}{2}\sum_{m,n}\Omega_{m,l}^{*}\tilde{\Delta}_{m}^{-1}\Omega_{m,n}
\Phi_{\mathcal{E}_{j}s_{n}}\nonumber\\
&+i\frac{g}{2}\sum_{m}\left(\tilde{\Delta}_{m}^{-1}\Omega_{m,l}^{*}
  \Phi_{\mathcal{E}_{j}\mathcal{E}_{m}}+\tilde{\Delta}_{j}^{-1}\Omega_{j,m}
  \Phi_{s_{m}s_{l}}\right)\,,
\label{eq:es-t}\\
\partial_{t}\Phi_{s_{j}s_{l}} = &
i\frac{g}{2}\sum_{m}\tilde{\Delta}_{m}^{-1}(\Omega_{m,j}^{*}\Phi_{\mathcal{E}_{m}s_{l}}
+\Omega_{m,l}^{*}\Phi_{s_{j}\mathcal{E}_{m}})
+\frac{i}{2}\sum_{m,n}\tilde{\Delta}_{m}^{-1}\Omega_{m,n}(\Omega_{mj}^{*}
\Phi_{s_{n}s_{l}}+\Omega_{m,l}^{*}\Phi_{s_{j}s_{n}})\nonumber\\
&-iV(z'-z)\Phi_{s_{j}s_{l}}\,.
\label{eq:ss-t}
\end{align}
The envelopes of the input probe fields are assumed to be constant most of the
time, so the boundary conditions for Eqs.~(\ref{eq:ee-t})--(\ref{eq:ss-t}) are
time independent. Hence one can consider the steady state of the fields by
dropping the time derivatives in Eqs.~(\ref{eq:ee-t})--(\ref{eq:ss-t}).  To
simplify the equations it is convenient to introduce the following combinations
of two-excitation wave functions:
\begin{align}
\Phi_{j,l}^{\mathcal{E}s}(z,z') & =
\frac{1}{g}\sum_{m}\Omega_{lm}\Phi_{\mathcal{E}_{j}s_{m}}(z,z')\,,\\
\Phi_{j,l}^{s\mathcal{E}}(z,z') & =
\frac{1}{g}\sum_{m}\Omega_{jm}\Phi_{s_{m}\mathcal{E}_{l}}(z,z')\,,\\
\Phi_{j,l}^{ss}(z,z') & =
\frac{1}{g^{2}}\sum_{m,n}\Omega_{jm}\Omega_{ln}\Phi_{s_{m}s_{n}}(z,z')\,.
\end{align}
Calling on Eqs.~(\ref{eq:ee-t})--(\ref{eq:ss-t}), one arrives at the steady
state equations for the combined wave functions $\Phi_{j,l}^{\mathcal{E}s}$,
$\Phi_{j,l}^{s\mathcal{E}}$ and $\Phi_{j,l}^{ss}$:
\begin{align}
0 & =-c(\partial_{z}+\partial_{z'})\Phi_{\mathcal{E}_{j}\mathcal{E}_{l}}
+i\frac{g^{2}}{2}\left((\tilde{\Delta}_{j}^{-1}+\tilde{\Delta}_{l}^{-1})
  \Phi_{\mathcal{E}_{j}\mathcal{E}_{l}}+\tilde{\Delta}_{l}^{-1}
  \Phi_{j,l}^{\mathcal{E}s}+\tilde{\Delta}_{j}^{-1}\Phi_{j,l}^{s\mathcal{E}}\right)\,,
\label{eq:ee-2}\\
0 & =-c\partial_{z}\Phi_{j,l}^{\mathcal{E}s}+i\frac{g^{2}}{2}
\tilde{\Delta}_{j}^{-1}(\Phi_{j,l}^{\mathcal{E}s}+\Phi_{j,l}^{ss})
+i\frac{g^{2}}{2}\sum_{m}\tilde{\Delta}_{m}^{-1}
\frac{v_{l,m}}{c}(\Phi_{j,m}^{\mathcal{E}s}+\Phi_{\mathcal{E}_{j}\mathcal{E}_{m}})\,,
\label{eq:es-2}\\
0 & =-c\partial_{z'}\Phi_{j,l}^{s\mathcal{E}}+
i\frac{g^{2}}{2}\tilde{\Delta}_{l}^{-1}(\Phi_{j,l}^{s\mathcal{E}}+
\Phi_{j,l}^{ss})+i\frac{g^{2}}{2}\sum_{m}\tilde{\Delta}_{m}^{-1}
\frac{v_{j,m}}{c}(\Phi_{m,l}^{s\mathcal{E}}+
\Phi_{\mathcal{E}_{m}\mathcal{E}_{l}})\,,
\label{eq:se-2}\\
0 & =\sum_{m}\tilde{\Delta}_{m}^{-1}\left(v_{j,m}(\Phi_{m,l}^{\mathcal{E}s}
  +\Phi_{m,l}^{ss})+v_{l,m}(\Phi_{j,m}^{s\mathcal{E}}+\Phi_{j,m}^{ss})\right)-
\frac{2c}{g^{2}}V(z'-z)\Phi_{j,l}^{ss}\,,
\label{eq:ss-2}
\end{align}
Here the Rabi frequencies enter only via the matrix elements $v_{j,l}$ of the
the group velocity matrix, defined by Eq.~(\ref{eq:v-matr}).  Numerical
solution of Eqs.~(\ref{eq:ee-2})--(\ref{eq:ss-2}) for various values of the
one-photon detunings $\Delta_{1}$ and $\Delta_{2}$ shows that the wave function
$\Phi_{\mathcal{E}_{j}\mathcal{E}_{l}}$ can be approximated as
\begin{equation}
\Phi_{\mathcal{E}_{j}\mathcal{E}_{l}}=-\frac{1}{2}(\Phi_{j,l}^{\mathcal{E}s}
+\Phi_{j,l}^{s\mathcal{E}})\,.\label{eq:ee-es-plus}
\end{equation}
The last term in Eqs.~(\ref{eq:es-2}) and (\ref{eq:se-2}) contains the ratio of
the group velocity to the speed of light $v/c$. Under the EIT conditions the
group velocity $v$ is much smaller than the speed of light, $v/c\ll1$, so we
neglect the last term in Eqs.~(\ref{eq:es-2}) and (\ref{eq:se-2}), giving
\begin{align}
0 & =-c\partial_{z}\Phi_{j,l}^{\mathcal{E}s}+i\frac{g^{2}}{2}
\tilde{\Delta}_{j}^{-1}(\Phi_{j,l}^{\mathcal{E}s}+\Phi_{j,l}^{ss})\,,
\label{eq:es-3}\\
0 & =-c\partial_{z'}\Phi_{j,l}^{s\mathcal{E}}+
i\frac{g^{2}}{2}\tilde{\Delta}_{l}^{-1}(\Phi_{j,l}^{s\mathcal{E}}+
\Phi_{j,l}^{ss})\,.
\label{eq:se-3}
\end{align}
To get a closed equation we need to solve Eq.~(\ref{eq:ss-2}) to express the
wave functions $\Phi_{j,l}^{ss}$ via $\Phi_{j,l}^{\mathcal{E}s}$ and
$\Phi_{j,l}^{s\mathcal{E}}$. As it is shown in the
Appendix~\ref{sec:spec-solution}, the solution of Eq.~(\ref{eq:ss-2}) can be
written in the following form:
\begin{multline}
\Phi_{j,l}^{ss}+\Phi_{j,l}^{\mathcal{E}s+}=\frac{1}{\tilde{\Delta}_{1}^{-1}v_{1,1}+
  \tilde{\Delta}_{2}^{-1}v_{2,2}-
  \frac{2c}{g^{2}}V(z'-z)}\left(\sum_{m}\tilde{\Delta}_{m}^{-1}(v_{l,m}
  \Phi_{j,m}^{\mathcal{E}s-}-v_{j,m}\Phi_{m,l}^{\mathcal{E}s-})\right.\\
\left.-\frac{2c}{g^{2}}V(z'-z)\sum_{m,n}A_{j,l;m,n}\Phi_{m,n}^{\mathcal{E}s+}\right)\,,
\label{eq:ss-sol}
\end{multline}
where
\begin{align}
\Phi_{j,l}^{\mathcal{E}s+} & =\frac{1}{2}(\Phi_{j,l}^{\mathcal{E}s}+
\Phi_{j,l}^{s\mathcal{E}})\,,\\
\Phi_{j,l}^{\mathcal{E}s-} & =\frac{1}{2}(\Phi_{j,l}^{\mathcal{E}s}-
\Phi_{j,l}^{s\mathcal{E}})\,.
\end{align}
Note that the wave functions $\Phi_{j,l}^{\mathcal{E}s+}$ and
$\Phi_{j,l}^{\mathcal{E}s-}$ represent a generalization of the symmetric and
antisymmetric combinations of two-excitation wave functions used in
Ref.~\cite{Peyronel2012} for a simpler ladder scheme. The coefficients
$A_{j,l;m,n}$ in Eq.~(\ref{eq:ss-sol}) are obtained from the solution of linear
equations (\ref{eq:app-eq-y}) written in the form (\ref{eq:app-y-form}). We do
not present analytical expressions for the coefficients $A_{j,l;m,n}$
explicitly, since these expressions are long and not informative.

\section{Equal one-photon detunings\label{sec:equal-detunings}}

\subsection{Closed equation for the two-photon wave function}

In this Section we will consider the case of equal one-photon detunings,
$\Delta_{1}=\Delta_{2}\equiv\Delta$. In this situation one can obtain a closed
approximate equation for the two-photon wave function. To do this we will
employ of the center of mass and relative coordinates
\begin{equation}
R=\frac{1}{2}(z+z')\,,\qquad r=z-z'\,.\label{eq:R,r}
\end{equation}
Note, that $R$ and $r$ form a pair of mutually indepent coordinates that can be
used instead of the initial coordinates $z$ and $z'$. Using the center of mass
and relative coordinates, Eqs.~(\ref{eq:es-3}), (\ref{eq:se-3}) for the
combined wave functions $\Phi_{j,l}^{\mathcal{E}s+}$,
$\Phi_{j,l}^{\mathcal{E}s-}$ become
\begin{align}
0 & =-\frac{c}{2}\partial_{R}\Phi_{j,l}^{\mathcal{E}s+}-
c\partial_{r}\Phi_{j,l}^{\mathcal{E}s-}+
i\frac{g^{2}}{2}\tilde{\Delta}^{-1}(\Phi_{j,l}^{\mathcal{E}s+}+\Phi_{j,l}^{ss})\,,
\label{eq:es-plus}\\
0 & =-\frac{c}{2}\partial_{R}\Phi_{j,l}^{\mathcal{E}s-}-
c\partial_{r}\Phi_{j,l}^{\mathcal{E}s+}+i\frac{g^{2}}{2}
\tilde{\Delta}^{-1}\Phi_{j,l}^{\mathcal{E}s-}\,.
\label{eq:es-minus}
\end{align}
The symmetric combination $\Phi_{j,l}^{\mathcal{E}s+}$ enters the last term of
equation (\ref{eq:ee-2}) and is thus directly coupled to
$\Phi_{\mathcal{E}_{j}\mathcal{E}_{l}}$. The antisymmetric combination
$\Phi_{j,l}^{\mathcal{E}s-}$ is decoupled from $\Phi_{j,l}^{ss},$ as one can
see in the equation (\ref{eq:es-minus}) for $\Phi_{j,l}^{\mathcal{E}s-}$ . We
next simplify Eqs.~(\ref{eq:es-plus})--(\ref{eq:es-minus}) using several
approximations, similar to the ones employed in
Refs.~\cite{Peyronel2012,Firstenberg2013}.  In Eq.~(\ref{eq:es-minus}) we
neglect the spatial derivative $\partial_{R}\Phi_{j,l}^{\mathcal{E}s-}$, thus
relating $\Phi_{j,l}^{\mathcal{E}s-}$ to $\Phi_{j,l}^{\mathcal{E}s+}$:
\begin{equation}
\Phi_{j,l}^{\mathcal{E}s-}=-2i\frac{c}{g^{2}}\tilde{\Delta}
\partial_{r}\Phi_{j,l}^{\mathcal{E}s+}\,.\label{eq:es-minus-es-plus}
\end{equation}
Inserting Eq.~(\ref{eq:es-minus-es-plus}) into Eq.~(\ref{eq:es-plus}) and using
Eq.~(\ref{eq:ss-sol}) together with Eq.~(\ref{eq:ee-es-plus}) one arrives at a
closed equation for the two-photon wave functions
\begin{multline}
i\partial_{R}\Phi_{\mathcal{E}_{j}\mathcal{E}_{l}}=
-4L_{\mathrm{abs}}\frac{\tilde{\Delta}}{\Gamma}
\partial_{r}^{2}\Phi_{\mathcal{E}_{j}\mathcal{E}_{l}}+
\frac{i}{\bar{v}-L_{\mathrm{abs}}\frac{\tilde{\Delta}}{\Gamma}V(r)}
\sum_{m}(v_{l,m}\partial_{r}\Phi_{\mathcal{E}_{j}\mathcal{E}_{m}}-
v_{j,m}\partial_{r}\Phi_{\mathcal{E}_{m}\mathcal{E}_{l}})\\
+\frac{V(r)}{\bar{v}-L_{\mathrm{abs}}\frac{\tilde{\Delta}}{\Gamma}V(r)}
\sum_{m,n}A_{jl,mn}\Phi_{\mathcal{E}_{m}\mathcal{E}_{n}}\,.
\label{eq:main-1}
\end{multline}
Here
\begin{equation}
\bar{v}=\frac{1}{2}(v_{1,1}+v_{2,2})
\end{equation}
is the average group velocity and 
\begin{equation}
L_{\mathrm{abs}}=\frac{L}{\alpha}
\end{equation}
is a resonant absorption length.

Equation (\ref{eq:main-1}), describing the propagation of two probe photons in
the medium, reveals several physical effects occurring during such a
propagation. Compared to a ladder scheme typically used for Rydberg EIT
\cite{Peyronel2012,Firstenberg2013}, the double tripod setup, considered in
this paper, has more tunable parameters.  There are two limiting cases of
Eq.~(\ref{eq:main-1}) corresponding to large and small separation $r$ between
two photons (compared to the Rydberg blockade radius). For separation distances
$r$ much larger than the blockade radius, the interaction potential $V(r)$
vanishes and Eq.~(\ref{eq:main-1}) reduces to
\begin{equation}
i\partial_{R}\Phi_{\mathcal{E}_{j}\mathcal{E}_{l}}=
-4L_{\mathrm{abs}}\left(\frac{2\Delta}{\Gamma}-
  i\right)\partial_{r}^{2}\Phi_{\mathcal{E}_{j}\mathcal{E}_{l}}+
\frac{i}{\bar{v}}\sum_{m}(v_{l,m}\partial_{r}\Phi_{\mathcal{E}_{j}\mathcal{E}_{m}}-
v_{j,m}\partial_{r}\Phi_{\mathcal{E}_{m}\mathcal{E}_{l}})\,.
\label{eq:equal-1}
\end{equation}

When the one-photon detuning is large compared to the rate of the excited state
delay, $\Delta\gg\Gamma$, Eq.~(\ref{eq:equal-1}) has the form of a
Schr\"odinger equation where the center of mass coordinate $R$ plays a role of
``time'' and the relative coordinate $r$ enters as a ``spatial coordinate''.
The first term on the right hand side of Eq.~(\ref{eq:equal-1}) represents the
kinetic energy corresponding to an effective mass
$\tilde{m}=\Gamma/(16L_{\mathrm{abs}}\Delta)$.  The effective mass can be
positive or negative, depending on the sign of the one-photon detuning
$\Delta_{1}$. This effective mass is the same as in the case of a simpler
ladder scheme for Rydberg EIT \cite{Peyronel2012,Firstenberg2013}.  If the
off-diagonal matrix elements of the group velocity matrix $v_{j,l}$ are
nonzero, the second term on the on the right hand side of
Eq.~(\ref{eq:equal-1}) couples the linear momentum, represented by the
firs-order derivative $\partial_{r}$, with two-photon wave functions
$\Phi_{\mathcal{E}_{j}\mathcal{E}_{l}}$ having different indices. The wave
functions $\Phi_{\mathcal{E}_{j}\mathcal{E}_{l}}$ can be interpreted as a
four-component spinor, thus the last term of Eq.~(\ref{eq:equal-1}) represents
\emph{spin-orbit coupling for the photons}.

In the opposite case corresponding to small $r$, the potential $V(r)$ becomes
large, $V(r)\gg\Omega_{j,l}^{2}/\Delta_{j}$, so the propagation equation
(\ref{eq:main-1}) takes the form
\begin{equation}
i\partial_{R}\Phi_{\mathcal{E}_{j}\mathcal{E}_{l}}=
-4L_{\mathrm{abs}}\left(\frac{2\Delta}{\Gamma}-i\right)
\partial_{r}^{2}\Phi_{\mathcal{E}_{j}\mathcal{E}_{l}}-
\frac{1}{L_{\mathrm{abs}}}\frac{1}{\frac{2\Delta}{\Gamma}-i}
\Phi_{\mathcal{E}_{j}\mathcal{E}_{l}}\,.
\label{eq:equal-2}
\end{equation}
Here we have used the property $A_{j,l;m,n}\rightarrow\delta_{j,m}\delta_{l,n}$
when $V(r)\rightarrow\infty$ (see Appendix~\ref{sec:spec-solution}). When
one-photon detuning is much larger than the decay rate of the intermediate
levels, $\Delta\gg\Gamma$, the last term on the right-hand side of
Eq.~(\ref{eq:equal-2}) represents an effective potential for the photons
$\mathcal{V}(r)=\Gamma/(2L_{\mathrm{abs}}\Delta)$.  The sign of the effective
potential is the same as the sign of one-photon detuning $\Delta$.  Since the
effective mass also changes the sign, the effective force is always attractive
\cite{Firstenberg2013}.

We can conclude that for equal and large one-photon detunings exceeding the
decay rate $\Gamma$, the propagation equation (\ref{eq:main-1}) for the
two-photon wave functions describes spin-orbit coupling at large distances
between photons together with an effective attractive interaction at small
distances. The spin-orbit coupling is a new feature of the double-tripod scheme
that is not present in a simpler ladder scheme. Note, that the spin-orbit term
decreases as $V(r)^{-1}$ and becomes small compared to other terms in the
equation at separation between photons satisfying the condition
$V(r)\gg\Omega_{j,l}^{2}/\Delta$.  These are distances smaller than the
blockade radius $r_{\mathrm{b}}$ given by Eq.~(\ref{eq:rb}).

When one-photon detuning $\Delta$ is small compared to the decay rate $\Gamma$
or zero, the first term on the right hand side of the propagation equation
(\ref{eq:main-1}) becomes imaginary and the equation acquires the form of a
diffusion equation. The diffusion term represents spreading out of the wave
packet of slow light caused by the non-adiabatic losses due to the deviation
from the EIT central frequency. The last term of Eq.~(\ref{eq:equal-2})
describes the absorption of the photons in the case when the distance between
the photons is small, representing the Rydberg blockade effect
\cite{Gorshkov2011,Peyronel2012}.

\subsection{Second-order correlation functions}

\begin{figure}
\includegraphics[width=0.33\textwidth]{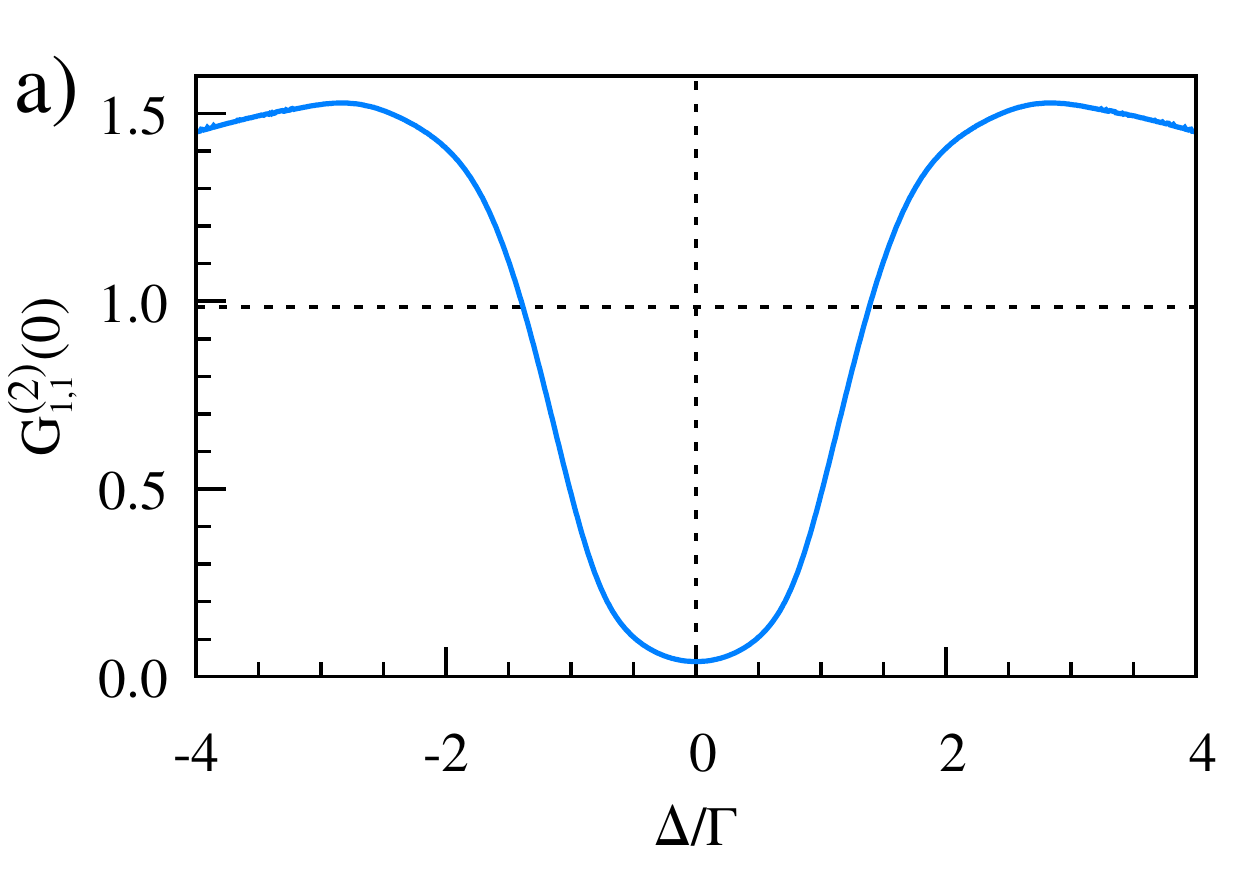}\includegraphics[width=0.33\textwidth]{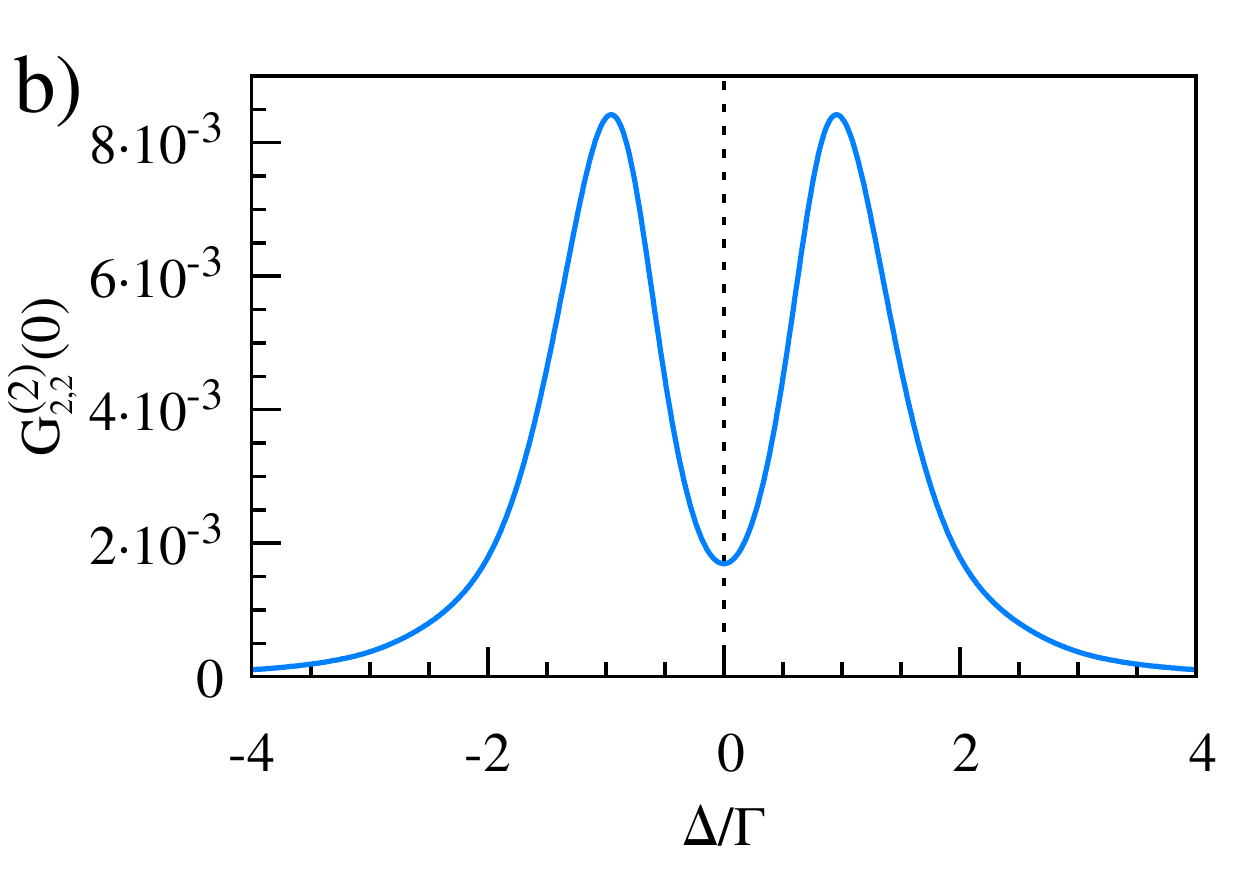}\includegraphics[width=0.33\textwidth]{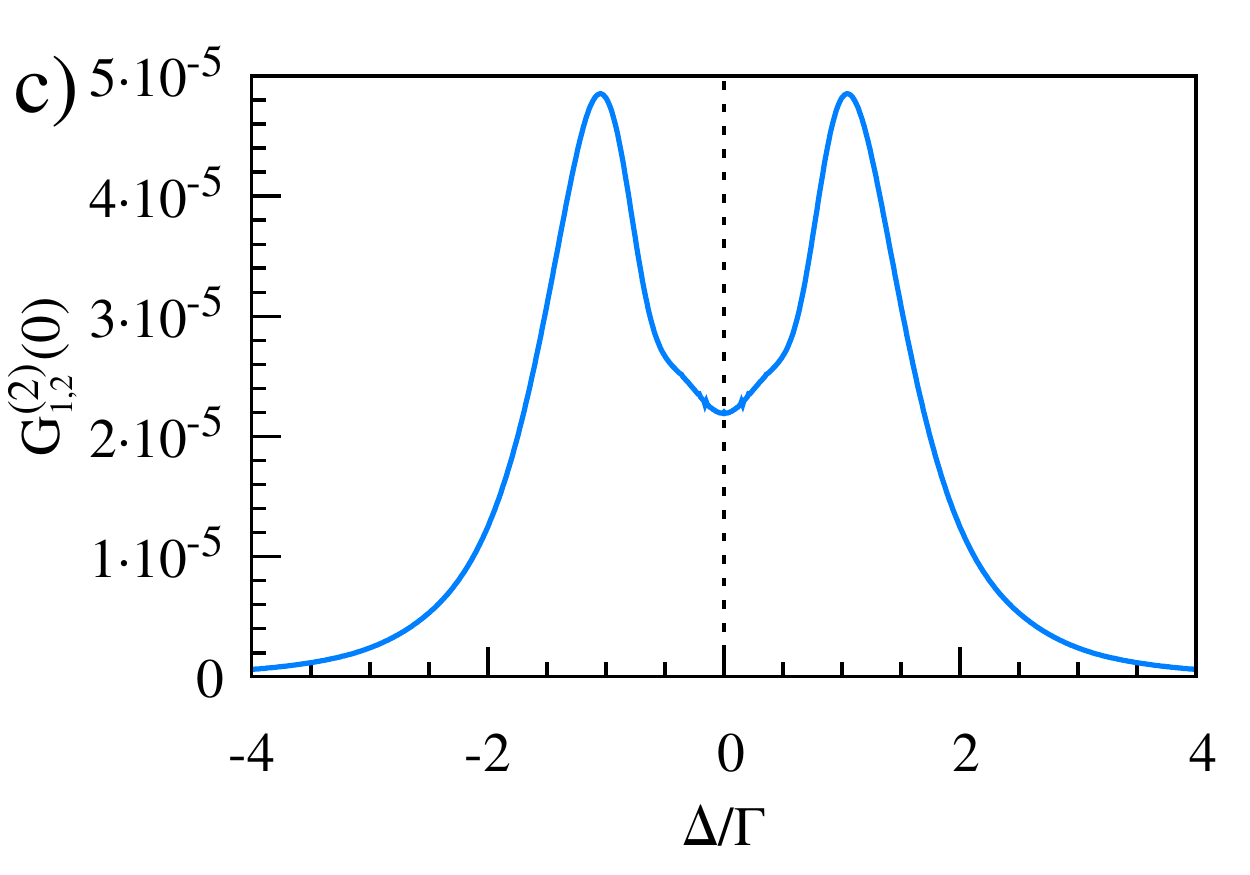}
\caption{The second-order correlation functions of the probe beams in the
  output normalized to the intensity of the input pulse. The correlation
  functions are calculated using Eq.~(\ref{eq:correlation-def}), where the wave
  functions are obtained by solving Eqs.~($\ref{eq:equal-2}$)
  (\ref{eq:equal-3}).  The parameters used in numerical solution are: optical
  density $\alpha=30$, Rydberg blockade radius
  $r_{\mathrm{b}}/L_{\mathrm{abs}}=0.4$, and ratio of non-diagonal elements of
  the group velocity matrix to the diagonal elements $v_{1,2}/v_{1,1}=1/2$.}
\label{fig:equal-delta}
\end{figure}

As an example, let us consider a situation where the diagonal matrix elements
of the group velocity matrix are equal, $v_{1,1}=v_{2,2}$ and off-diagonal
elements are real-valued, giving $v_{1,2}=v_{2,1}$.  This can be the case when
all Rabi frequencies of the control fields are real-valued and satisfy the
conditions $\Omega_{2,2}=\Omega_{1,1},\Omega_{2,1}=\Omega_{1,2}$.  In this
situation the group velocity matrix 
\begin{equation}
v=\frac{L_{\mathrm{abs}}}{\Gamma}\left(\begin{array}{cc}
\Omega_{11}^{2}+\Omega_{12}^{2} & 2\Omega_{11}\Omega_{12}\\
2\Omega_{11}\Omega_{12} & \Omega_{11}^{2}+\Omega_{12}^{2}
\end{array}\right)
\end{equation}
has eigenvalues
\begin{equation}
v_{\pm}=\frac{L_{\mathrm{abs}}}{\Gamma}(\Omega_{11}\pm\Omega_{12})^{2}\,.
\end{equation}
Another possibility to have such a velocity matrix is equal absolute values of
the Rabi frequencies, $\Omega_{j,l}=\Omega e^{iS_{j,l}}$, together with the
phases $S_{1,1}=S_{2,2}=0$ and $S_{1,2}=S_{2,1}=S\neq0$.  The group velocity
matrix then is
\begin{equation}
\hat{v}=\frac{2L_{\mathrm{abs}}\Omega^{2}}{\Gamma}\left(\begin{array}{cc}
1 & \cos S\\
\cos S & 1
\end{array}\right)
\end{equation}
and has eigenvalues
\begin{equation}
v_{\pm}=\frac{2L_{\mathrm{abs}}\Omega^{2}}{\Gamma}(1\pm\cos S)\,.
\end{equation}
For the group velocity matrix having equal diagonal and off-diagonal elements,
Eq.~(\ref{eq:equal-1}) for the distances $r$ much larger than the blockade
radius becomes
\begin{equation}
i\partial_{R}\Phi_{\mathcal{E}_{j}\mathcal{E}_{l}}=
-4L_{\mathrm{abs}}\left(\frac{2\Delta}{\Gamma}-i\right)\partial_{r}^{2}
\Phi_{\mathcal{E}_{j}\mathcal{E}_{l}}+i\frac{v_{1,2}}{v_{1,1}}
\sum_{m}\partial_{r}(\Phi_{\mathcal{E}_{j}\mathcal{E}_{m}}-
\Phi_{\mathcal{E}_{m}\mathcal{E}_{l}})\,.
\label{eq:equal-3}
\end{equation}

We will investigate the situation when only the first probe beam with the
amplitude $a$ is incident on the atom cloud. Due to the non-diagonal matrix
elements of the group velocity matrix and the presence of the atom-atom
interaction $V$, outside of the atomic medium not only the first probe beam but
also the second beam will be present. Since there are two probe beams, one can
consider four second-order correlation functions that describe correlations
between the light from the same beam as well as correlations between different
beams. We will use the second-order correlation functions normalized to the
intensity of the incident probe beam,
\begin{equation}
G_{j,l}^{(2)}(0)=\frac{1}{a^{4}}|\Phi_{\mathcal{E}_{j}\mathcal{E}_{l}}(R=L,r=0)|^{2}\,.
\label{eq:correlation-def}
\end{equation}
Note, that the second-order correlation functions $G_{j,l}^{(2)}$ do not depend
on the amplitude of the incident beam $a$, because
$\Phi_{\mathcal{E}_{j}\mathcal{E}_{l}}$ are proportional to $a^{2}$.  The
situation when the first probe beam is incident on the atom cloud corresponds
to the boundary conditions for two-photon wave functions
$\Phi_{\mathcal{E}_{1}\mathcal{E}_{1}}(z=0,z')=
\Phi_{\mathcal{E}_{1}\mathcal{E}_{1}}(z,z'=0)=a^{2}$
and
$\Phi_{\mathcal{E}_{j}\mathcal{E}_{l}}(z=0,z')=
\Phi_{\mathcal{E}_{j}\mathcal{E}_{l}}(z,z'=0)=0$
when $j\neq1$, $l\neq1$. To simplify the solution we make an approximation by
assuming that the boundary conditions are
$\Phi_{\mathcal{E}_{1}\mathcal{E}_{1}}(R=0,r)=
\Phi_{\mathcal{E}_{1}\mathcal{E}_{1}}(R,r=\pm\alpha)=a^{2}$
and
$\Phi_{\mathcal{E}_{j}\mathcal{E}_{l}}(R=0,r)=
\Phi_{\mathcal{E}_{j}\mathcal{E}_{l}}(R,r=\pm\alpha)=0$
when $j\neq1$, $l\neq1$ \cite{Firstenberg2013}. In addition, we approximate the
atom-atom interaction potential $V(r)$ as having a large value when
$|r|<\bar{r}_{\mathrm{B}}$ and zero when $|r|>\bar{r}_{\mathrm{B}}$, where
\begin{equation}
\bar{r}_{\mathrm{B}}=r_{\mathrm{b}}\left(\left(\frac{2\Delta}{\Gamma}\right)^{2}
  +1\right)^{\frac{1}{12}}
\end{equation}
is the blockade radius increased due to the presence of nonzero one-photon
detuning $\Delta$. Thus the propagation equation for two-photon wave functions
is given by Eq.~($\ref{eq:equal-2}$) for $|r|<\bar{r}_{\mathrm{B}}$ and by
Eq.~(\ref{eq:equal-3}) for $|r|>\bar{r}_{\mathrm{B}}$.

For numerical solution we take the optical depth $\alpha=30$, the ratio of the
blockade radius to the absorption length $r_{\mathrm{b}}/L_{\mathrm{abs}}=0.4$
and the ratio of non-diagonal elements of the group velocity matrix to the
diagonal elements $v_{1,2}/v_{1,1}=1/2$. The resulting dependence of
second-order correlation functions $G_{j,l}^{(2)}(0)$ on the one-photon
detuning $\Delta$, obtained by solving Eqs.~($\ref{eq:equal-2}$),
(\ref{eq:equal-3}), is shown in Fig.~\ref{fig:equal-delta}. As one can see in
Fig.~\ref{fig:equal-delta}a, the second-order correlation function
$G_{1,1}^{(2)}(0)$ is much smaller than $1$ when one-photon detuning $|\Delta|$
is smaller that the decay rate $\Gamma$, indicating the anti-bunching of
photons which arises due to the destruction of the EIT because of the Rydberg
blockade. On the other hand, for large values of the one-photon detuning the
second-order correlation function $G_{1,1}^{(2)}(0)$ is larger than $1$ and the
photons are bunched.  This bunching is a result of an effective attractive
interaction between the photons \cite{Firstenberg2013,Firstenberg2016}. The
main difference of the double-tripod scheme from the simple ladder scheme lies
in the creation of the second probe beam, which is indicated by nonzero
second-order correlation functions $G_{2,2}^{(2)}(0)$ and $G_{1,2}^{(2)}(0)$,
shown in Figs.~\ref{fig:equal-delta}b and \ref{fig:equal-delta}c.  Although
transfer of the photons from the first to the second probe beams is possible in
a double-tripod setup without atom-atom interactions, to do so a nonzero
two-photon detuning is needed \cite{Ruseckas2013,Lee2014}.  Here we considered
the situation where two-photon detuning is zero and the envelope of the input
probe field is constant most of the time. Therefore the appearance of the
second probe beam is caused by the interactions between Rydberg atoms. As can
be seen in Fig.~\ref{fig:equal-delta}b, there is an optimal value of the
one-photon detuning $|\Delta|$, where the intensity of the created second probe
beam is the largest.  This is caused by an interplay of two effects: for small
one-photon detuning $|\Delta|<\Gamma$ absorption of photons takes places due to
the Rydberg blockade, whereas the efficiency of the transfer of photons from
the first to the second probe beams decreases with increasing $|\Delta|$.

\section{Opposite signs of one-photon detunings\label{sec:opposite-signs}}

\begin{figure}
\includegraphics[width=0.4\textwidth]{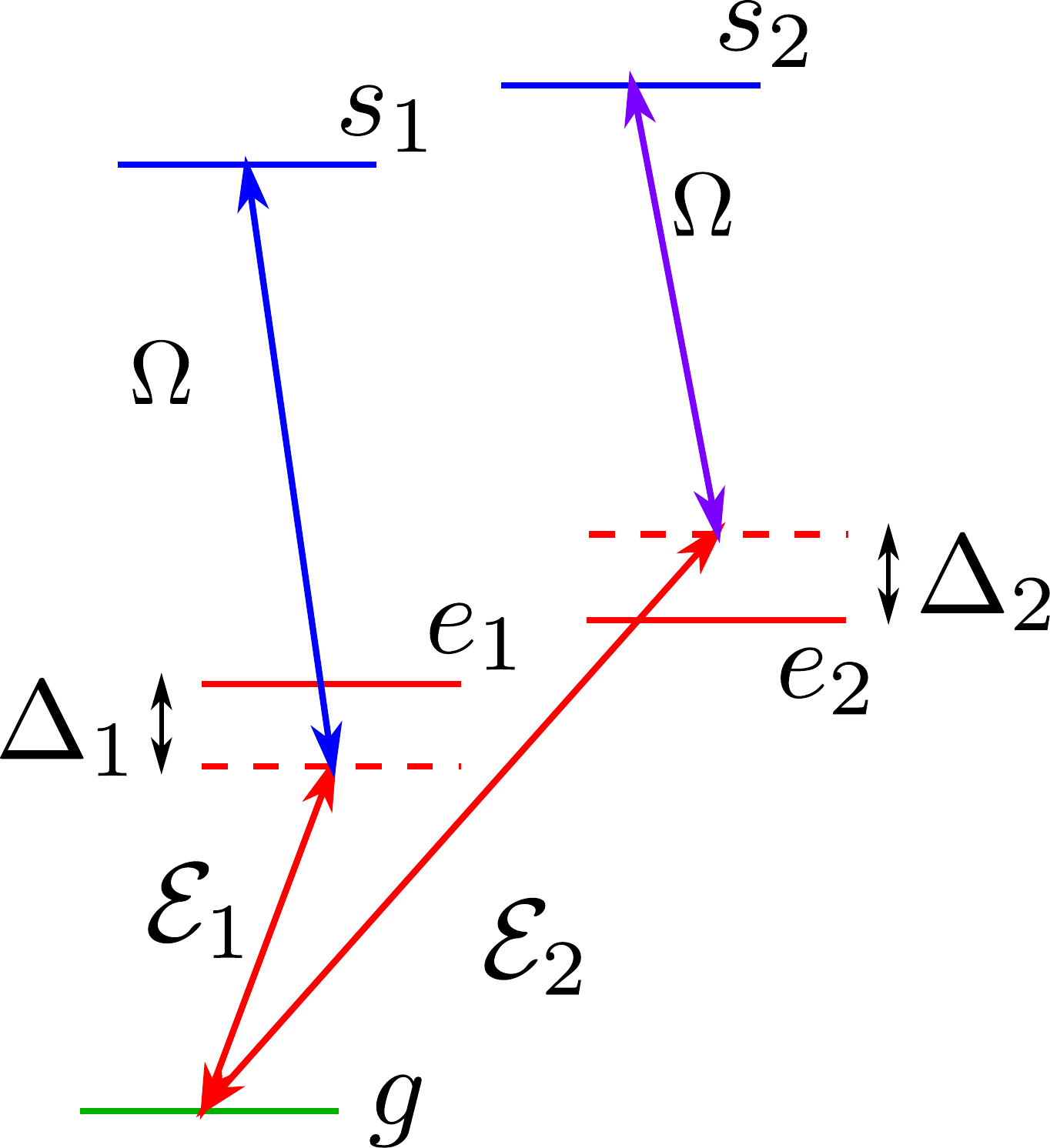}
\caption{Double ladder atom-light coupling scheme involving the ground level
  $g$, the Rydberg levels $s_{1}$ and $s_{2}$, and the intermediate excited
  levels $e_{1}$ and $e_{2}$. Two weak probe beams with the amplitudes
  $\mathcal{E}_{j}$ connect the ground level $g$ to the intermediate levels
  $e_{j}$. Two strong control beam with the Rabi frequencies $\Omega$ connect
  the intermediate levels $e_{j}$ to the Rydberg levels $s_{j}$.}
\label{fig:two-lambdas}
\end{figure}

In the case when one-photon detunings $\Delta_{1}$ and $\Delta_{2}$ are
different, numerical solution of Eqs.~(\ref{eq:es-3})--(\ref{eq:ss-sol}) shows
that the approximation made in deriving Eq.~(\ref{eq:es-minus-es-plus}) is no
longer valid. Thus one cannot obtain a single approximate equation, similar to
Eq.~(\ref{eq:main-1}) describing the propagation of probe fields in the case of
different one-photon detunings. To investigate the consequence of different
detunings in this Section we will consider a simpler double ladder atom-light
coupling scheme, shown in Fig.~\ref{fig:two-lambdas}.  This scheme is a
particular case of the double tripod scheme with control fields $\Omega_{12}$
and $\Omega_{21}$ equal to zero. A consequence of this simplification is that
the conversion of the photons between the different probe fields does not
occur. In addition we will take the Rabi frequencies of the two remaining
control fields to be equal: $\Omega_{11}=\Omega_{22}\equiv\Omega$. The group
velocity matrix (\ref{eq:v-matr}) for this situation becomes diagonal, with the
diagonal elements equal to $v=c|\Omega|^{2}/g^{2}$.

For the double ladder scheme, the expression (\ref{eq:ss-sol}) for the wave
function of two Rydberg excitations simplifies to 
\begin{equation}
\Phi_{j,l}^{ss}=\frac{1}{2}\frac{\tilde{\Delta}_{j}^{-1}\Phi_{j,l}^{\mathcal{E}s}+
  \tilde{\Delta}_{l}^{-1}\Phi_{j,l}^{s\mathcal{E}}}{\tilde{V}(z'-z)-
  \frac{1}{2}(\tilde{\Delta}_{j}^{-1}+\tilde{\Delta}_{l}^{-1})}\,,
\label{eq:ss-opposite}
\end{equation}
where
\begin{equation}
\tilde{V}(r)=\frac{1}{|\Omega|^{2}}V(r)\,.
\end{equation}
Inserting Eq.~(\ref{eq:ss-opposite}) into Eqs.~(\ref{eq:es-3})--(\ref{eq:se-3})
we obtain a closed pair of equations
\begin{align}
c\partial_{z}\Phi_{j,l}^{\mathcal{E}s} & =\frac{i}{2}g^{2}\tilde{\Delta}_{j}^{-1}
\frac{\tilde{V}(r)\Phi_{j,l}^{\mathcal{E}s}+
  \frac{1}{2}\tilde{\Delta}_{l}^{-1}(\Phi_{j,l}^{s\mathcal{E}}-
  \Phi_{j,l}^{\mathcal{E}s})}{\tilde{V}(r)-\frac{1}{2}(\tilde{\Delta}_{j}^{-1}+
  \tilde{\Delta}_{l}^{-1})}\,,\label{eq:es-4}\\
c\partial_{z'}\Phi_{j,l}^{s\mathcal{E}} & =
\frac{i}{2}g^{2}\tilde{\Delta}_{l}^{-1}\frac{\tilde{V}(r)\Phi_{j,l}^{s\mathcal{E}}+
  \frac{1}{2}\tilde{\Delta}_{j}^{-1}(\Phi_{j,l}^{\mathcal{E}s}-
  \Phi_{j,l}^{s\mathcal{E}})}{\tilde{V}(r)-\frac{1}{2}(\tilde{\Delta}_{j}^{-1}+
  \tilde{\Delta}_{l}^{-1})}\,.\label{eq:se-4}
\end{align}
The two-photon wave function $\Phi_{\mathcal{E}_{j}\mathcal{E}_{l}}$ can be
calculated from the solutions of Eqs.~(\ref{eq:es-4}), (\ref{eq:se-4}) using
Eq.~(\ref{eq:ee-es-plus}).

\begin{figure}
\includegraphics[width=0.45\textwidth]{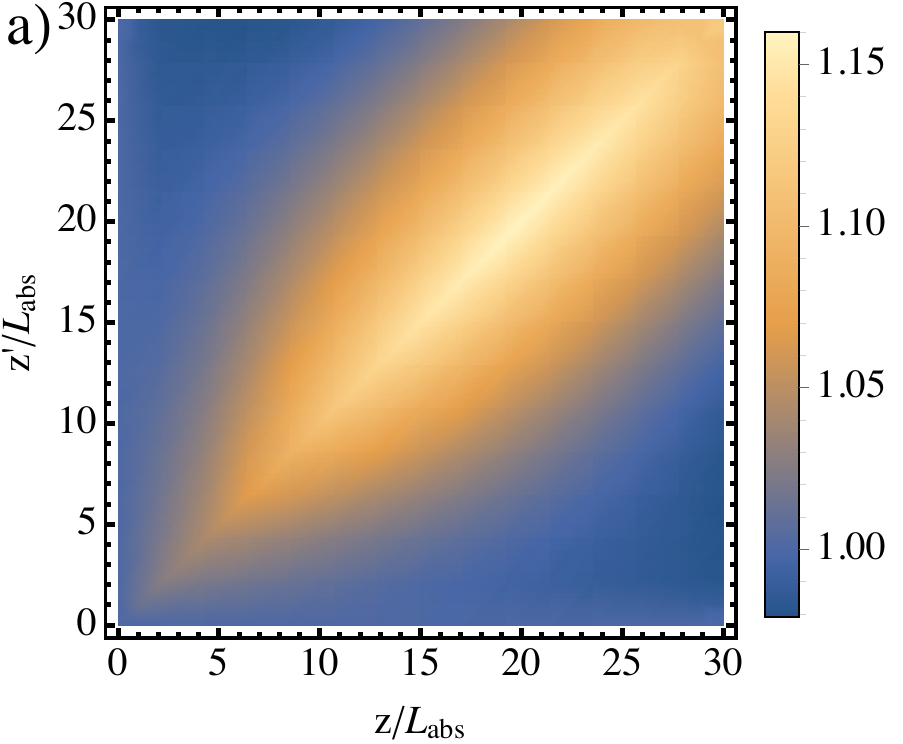}\includegraphics[width=0.45\textwidth]{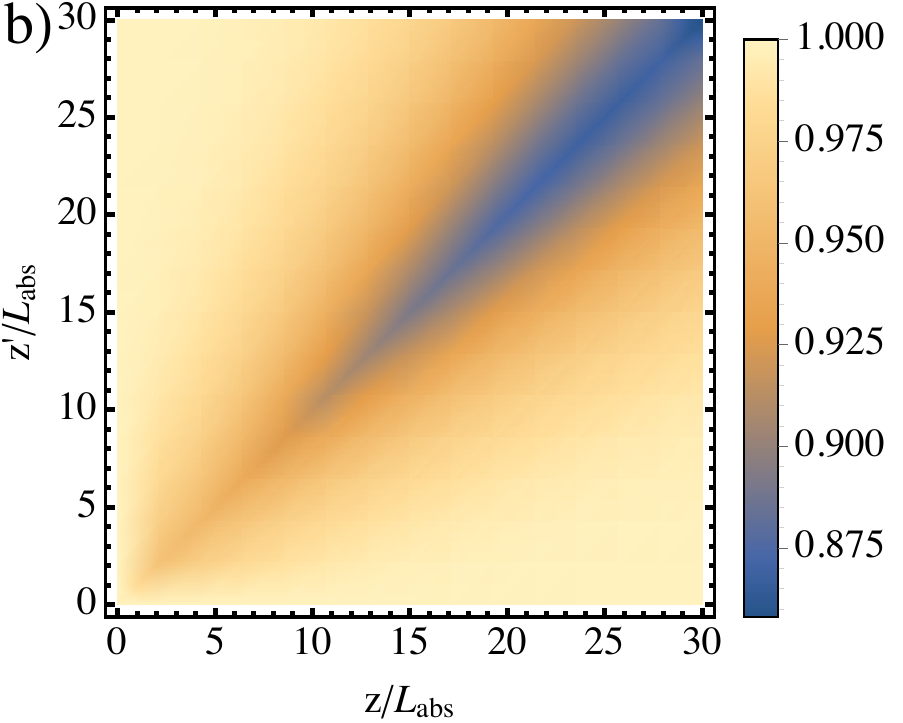}
\caption{Square of the absolute value of the two-photon wave function a)
  $|\Phi_{\mathcal{E}_{1}\mathcal{E}_{1}}(z,z')|^{2}$ and b)
  $|\Phi_{\mathcal{E}_{1}\mathcal{E}_{2}}(z,z')|^{2}$, when one-photon
  detunings have opposite signs. The wave functions are calculated using
  Eqs.~(\ref{eq:es-4}), (\ref{eq:se-4}) and (\ref{eq:ee-2}).  The parameters
  used in numerical solution are: optical density $\alpha=30$, Rydberg blockade
  radius $r_{\mathrm{b}}/L_{\mathrm{abs}}=0.4$, one-photon detunings
  $\Delta_{1}/\Gamma=-\Delta_{2}/\Gamma=2.5$.}
\label{fig:opposite-delta}
\end{figure}

Let us consider the situation when the one-photon detunings have opposite
signs, $\Delta_{2}=-\Delta_{1}$, and both the first and second probe beams with
the equal amplitudes are incident on the atom cloud. According to approximate
Eqs.~(\ref{eq:es-4}), (\ref{eq:se-4}), the two-photon wave function when both
photons are from the same probe beam ($j=l$) behaves similarly as in an single
ladder atom-light coupling scheme.  This can be seen in
Fig.~\ref{fig:opposite-delta}a, where the two-photon wave function has
amplitude larger than $1$ along the diagonal line $z=z'$, indicating the photon
bunching occuring when one-photon detuning is large, $|\Delta|>\Gamma$. More
interesting is the behavior of the two-photon wave function when the photons
are from different probe beams. Dependence of the square of the absolute value
of the two-photon wave function $\Phi_{\mathcal{E}_{1}\mathcal{E}_{2}}(z,z')$
on the positions $z$ and $z'$, obtained by numerically solving
Eqs.~(\ref{eq:es-4}), (\ref{eq:se-4}) and (\ref{eq:ee-2}), is shown in
Fig.~\ref{fig:opposite-delta}b.  We assumed that both beams have equal
intensities, which corresponds to the boundary conditions
$\Phi_{\mathcal{E}_{1}\mathcal{E}_{2}}(0,z')=\Phi_{\mathcal{E}_{1}\mathcal{E}_{2}}(z,0)=1$
and $\Phi_{1,2}^{\mathcal{E}s}(0,z')=\Phi_{1,2}^{s\mathcal{E}}(z,0)=-1$.  The
sign ``-'' in the latter boundary conditions represents the fact that for a
single photon inside the medium the conditions for EIT hold and the dark-state
polariton forms. For numerical solution we take the atom-atom interaction
potential $V(r)=C_{6}/r_{\mathrm{b}}^{6}$, optical depth $\alpha=30$, the ratio
of the blockade radius to the absorption length
$r_{\mathrm{b}}/L_{\mathrm{abs}}=0.4$ and the one-photon detunings
$\Delta_{1}=-\Delta_{2}=2.5\Gamma$.

As one can see in Fig.~\ref{fig:opposite-delta}b, the two-photon wave function,
representing the correlation between the photons from different beams, has
amplitude smaller than $1$ along the diagonal line $z=z'$. This is in contrast
to the two-photon wave function for the photons from the same beam, which has
amplitude larger than $1$ along the diagonal when one-photon detuning is larger
than $\Gamma$ (see Fig.~\ref{fig:opposite-delta}a). However, the decrease of
the probability to find the photons from different probe beams close to each
other is not caused by photon absorption due to the Rydberg blockade which is
prevented by a large one-photon detuning. We can interpret the resulting
decrease of the second-order correlation function $G_{1,2}^{(2)}(0)$ as an
effective repulsion between the photons from different probe beams. Note, that
repulsive photon interaction can be realized in a single ladder system when the
Rabi frequency of the control field is large, $|\Omega|>|\Delta|$
\cite{Bienias2014}. In contrast to a single ladder system, in the double ladder
system with the one-photon detunings of opposite signs, the repulsive
interaction appears also for smaller Rabi frequencies $\Omega$.

\section{Summary and outlook\label{sec:conclusions}}

In summary, we have derived an approximate closed equation (\ref{eq:main-1})
describing the propagation of the spinor Rydberg polaritons in a double-tripod
setup with equal one-photon detunings. The equation shows that the dissipative
or dispersive nonlinearities are combined with a spin-orbit coupling for the
spinor polaritons in this system. As a consequence, the atom-atom interaction
can cause transfer of a photon from one probe beam to the other. In case where
the one-photon detunings have the opposite signs, the numerical solution of
Eqs.~(\ref{eq:es-4}), (\ref{eq:se-4}) and (\ref{eq:ee-2}) shows the effective
repulsion of the photons from different probe beams.

The proposed double tripod scheme can be experimentally implemented using
a cold gas of $^{87}\mathrm{Rb}$ atoms \cite{Peyronel2012}. The hyperfine ground
state $|5S_{1/2},F=2,m_{F}=2\rangle$ in which the atoms are prepared, serves as
the ground state $|g\rangle$ in our scheme.  As an intermediate excited states
$|e_{1}\rangle$ and $|e_{2}\rangle$ one can use a hyperfine excited states
$|5P_{3/2},F=3,m_{F}=3\rangle$ and $|5P_{3/2},F=3,m_{F}=1\rangle$.
Experimentally accessible values of the length of the atomic medium and the
optical density are, respectively, $L=1\,\mathrm{mm}$ and $\alpha=30$
\cite{Chiu2014,Hsiao2014}.  The Rydberg blockade radius $r_{\mathrm{b}}$,
depending on the principal quantum number of the Rydberg levels, can be of the
order of $13\,\mathrm{\mu m}$ \cite{Peyronel2012}.

In a future work, the description of the propagation of the spinor slow light in
the medium with strong atom-atom interactions should be extended from the two-body
description presented in this article, to a true many-body treatment of the
system. Furthermore, the photon-photon interaction can be increased by placing
an optical cavity around the atomic ensemble, resulting in the cavity-Rydberg
polaritons \cite{Parigi2012,Ningyuan2016,Boddeda2016}.  Thus another direction
of the research includes description of spinor Rydberg polaritons in a cavity.

\begin{acknowledgments}
The work was supported by the project TAP LLT-2/2016 of the Research Council of
Lithuania and the Ministry of Science and Technology of Taiwan under Grant
Nos.~106-2119-M-007-003 and 105-2923-M-007-002-MY3.  Authors also acknowledge
a support by the National Center for Theoretical Sciences, Taiwan.
\end{acknowledgments}

\appendix

\section{Solution of the equation (\ref{eq:ss-2})\label{sec:spec-solution}}

In this appendix we solve the system of four equations
\[
\sum_{m}\tilde{\Delta}_{m}^{-1}\left(v_{j,m}(\Phi_{m,l}^{\mathcal{E}s}+
  \Phi_{m,l}^{ss})+v_{l,m}(\Phi_{j,m}^{s\mathcal{E}}+
  \Phi_{j,m}^{ss})\right)-\frac{2c}{g^{2}}V(r)\Phi_{j,l}^{ss}=0
\]
for the four unknown quantities $\Phi_{j,l}^{ss}$. Here $r=z-z'$.  To separate
the solution into two parts let us introduce two new quantities $X_{j,l}$ and
$Y_{j,l}$ such that
\begin{equation}
\Phi_{j,l}^{ss}=X_{j,l}+Y_{j,l}-\Phi_{j,l}^{\mathcal{E}s+}\,,
\label{eq:app-x-y}
\end{equation}
where
\begin{equation}
\Phi_{j,l}^{\mathcal{E}s+}=\frac{1}{2}(\Phi_{j,l}^{\mathcal{E}s}+
\Phi_{j,l}^{s\mathcal{E}})
\end{equation}
is the symmetric combination of the wave functions. This leads to the equation
\begin{multline}
\sum_{m}\tilde{\Delta}_{m}^{-1}\left(v_{j,m}(X_{m,l}+Y_{m,l}+
  \Phi_{m,l}^{\mathcal{E}s-})+v_{l,m}(X_{j,m}+Y_{j,m}-
  \Phi_{j,m}^{\mathcal{E}s-})\right)\\
-\frac{2c}{g^{2}}V(r)(X_{j,l}+Y_{j,l}-\Phi_{j,l}^{\mathcal{E}s+})=0\,,
\end{multline}
where
\begin{equation}
\Phi_{j,l}^{\mathcal{E}s-}=\frac{1}{2}(\Phi_{j,l}^{\mathcal{E}s}-
\Phi_{j,l}^{s\mathcal{E}})
\end{equation}
is the antisymmetric combination. Since Eq.~(\ref{eq:app-x-y}) defines only the
sum of the quantities $X_{j,l}$ and $Y_{j,l}$, we have a freedom to impose an
additional condition on those quantities. We require for the quantities
$X_{j,l}$ to obey the equation
\begin{equation}
\sum_{m}\tilde{\Delta}_{m}^{-1}\left(v_{j,m}(X_{m,l}+
  \Phi_{m,l}^{\mathcal{E}s-})+v_{l,m}(X_{j,m}-\Phi_{j,m}^{\mathcal{E}s-})\right)-
\frac{2c}{g^{2}}V(r)X_{j,l}=0
\label{eq:app-eq-x}
\end{equation}
and for the quantities $Y_{j,l}$ to obey the equation
\begin{equation}
\sum_{m}\tilde{\Delta}_{m}^{-1}(v_{j,m}Y_{m,l}+
v_{l,m}Y_{j,m})-\frac{2c}{g^{2}}V(r)Y_{j,l}=
-\frac{2c}{g^{2}}V(r)\Phi_{j,l}^{\mathcal{E}s+}
\label{eq:app-eq-y}
\end{equation}
The solution of Eq.~(\ref{eq:app-eq-x}) has a short analytical expression
\begin{equation}
X_{j,l}=\frac{1}{\tilde{\Delta}_{1}^{-1}v_{1,1}+\tilde{\Delta}_{2}^{-1}v_{2,2}-
  \frac{2c}{g^{2}}V(r)}\sum_{m}\tilde{\Delta}_{m}^{-1}(v_{l,m}
\Phi_{j,m}^{\mathcal{E}s-}-v_{j,m}\Phi_{m,l}^{\mathcal{E}s-})\,.
\end{equation}
The analytical expression for the solution of Eq.~(\ref{eq:app-eq-y}) is
complicated. Since Eq.~(\ref{eq:app-eq-y}) is linear, the general form of the
solution can be written as a linear combination of the terms on the right hand
side of Eq.~(\ref{eq:app-eq-y}):
\begin{equation}
Y_{j,l}=-\frac{\frac{2c}{g^{2}}V(r)}{\tilde{\Delta}_{1}^{-1}v_{1,1}+
  \tilde{\Delta}_{2}^{-1}v_{2,2}-\frac{2c}{g^{2}}V(r)}
\sum_{m,n}A_{j,l;m,n}\Phi_{m,n}^{\mathcal{E}s+}\,,
\label{eq:app-y-form}
\end{equation}
where the coefficients $A_{j,l;m,n}$ can be obtained from the solution of
Eq.~(\ref{eq:app-eq-y}). In the limit of large interaction,
$V(r)\gg|\Omega_{j,l}|^2/|\tilde{\Delta}_j|$ , one can drop the terms in
Eq.~(\ref{eq:app-eq-y}) that are not proportional to $V(r)$ and obtain the
solution $Y_{j,l}\approx\Phi_{j,l}^{\mathcal{E}s+}$. Thus in this limit the
coefficients reduce to $A_{j,l;m,n}\approx\delta_{j,m}\delta_{l,n}$.

\end{document}